\documentclass[a4paper,11pt]{article}
\pdfoutput=1 

\usepackage{jheppub} 

\usepackage[T1]{fontenc} 

\usepackage{amsmath,amssymb,epsfig,ulem,color,slashed,multirow,wasysym}

\allowdisplaybreaks  

\def \beq{\begin{equation}}
\def \eeq{\end{equation}}
\def \bea{\begin{eqnarray}}
\def \eea{\end{eqnarray}}

\title{\boldmath Searching for $t \to c (u) h$ with dipole moments}

\author[a]{Martin Gorbahn}
\author[b,c]{and Ulrich Haisch}

\affiliation[a]{Department of Mathematical Sciences, University of Liverpool, \\ L69 3BX Liverpool, United Kingdom}

\affiliation[b]{Rudolf Peierls Centre for Theoretical Physics,
    University of Oxford, \\ OX1 3PN Oxford, United Kingdom}
    
\affiliation[c]{CERN, Theory Division, \\ CH-1211 Geneva 23, Switzerland}

\emailAdd{martin.gorbahn@liverpool.ac.uk}
\emailAdd{u.haisch1@physics.ox.ac.uk}

\abstract{A discovery of flavour-changing Higgs-boson decays would constitute an undeniable signal of new physics. We derive model-independent constraints on the $t c h$ and $t u h$ couplings that arise from the bounds on hadronic electric dipole moments. Comparisons  of the present and future sensitivities with both the direct LHC constraints  and the indirect limits  from $D$-meson physics are also presented.}
  
\preprint{LTH~1006}

\begin{document} 

\maketitle

\flushbottom

\section{Introduction}
\label{sec:introduction}

The recent discovery of a Higgs-like state by ATLAS  \cite{Aad:2012tfa} and CMS \cite{Chatrchyan:2012ufa} opens up the exciting possibility to search for  flavour-changing neutral current~(FCNC) interactions involving  the exchange of this new boson.  Within the quark sector of the Standard Model~(SM)  such transitions are absent at tree level and suppressed at  loop level both by the Glashow-Iliopoulos-Maiani mechanism and the small inter-generational mixing as  encoded in the Cabibbo-Kobayashi-Maskawa~(CKM) matrix.  The resulting SM branching ratios are largest  for top-quark decays, but even in this case remain  utterly small, amounting to ${\rm Br} \left ( t \to c  h \right ) \simeq 3  \cdot 10^{-15}$ and ${\rm Br} \left ( t \to u  h \right ) \simeq 2  \cdot 10^{-17}$ \cite{AguilarSaavedra:2004wm}~(see also \cite{Mele:1998ag}).  Finding evidence for FCNC Higgs-boson decays taking place at measurable rates, would hence inevitable imply physics beyond the SM~(BSM), presumably related to new dynamics close to the $\rm TeV$ scale.

As a matter of fact, various  well-motivated BSM  scenarios such as two Higgs doublet models \cite{Bejar:2000ub}, supersymmetric extensions of the SM \cite{Guasch:1999jp,Eilam:2001dh}, warped extra dimensions~\cite{Azatov:2009na,Casagrande:2010si} or models based on the idea of partial compositeness~\cite{Agashe:2009di} can give rise to additional contributions to the $t \to c(u) h$ rates that are orders of magnitude in excess of the SM expectations. The relevant interactions  can be parameterised by the following Lagrangian 
\beq \label{eq:L}
{\cal L} \supset -\sum_{q=c,u} \left ( Y_{tq} \, \bar t_L q_R \hspace{0.25mm} h + Y_{qt} \, \bar q_L t_R \hspace{0.25mm} h \right )+ {\rm h.c.} \,, 
\eeq
where the couplings $Y_{tq}$ and $Y_{qt}$ are in general complex, $L$ and $R$ indicate whether the quarks are left-handed or right-handed and $h$ is the physical Higgs-boson field. Irrespectively of the underlying dynamics, the above FCNC Higgs-boson couplings  can be probed directly at the LHC by measuring tree-level decays of the top quark \cite{CMSPASSUS-13-002,CMSPASHIG-13-034,Aad:2014dya,Craig:2012vj,Greljo:2014dka}  or indirectly through precision measurements of low-energy observables \cite{Harnik:2012pb} if these receive corrections from loops involving the top quark and the Higgs boson. 

The main goal of this paper is to derive and to compare  the direct and indirect constraints that apply in the case of the $t \to c (u) h$ transitions.  Our particular focus will thereby be on CP-violating observables  such as electric dipole moments (EDMs).  In the context of  lepton-flavour violation the contributions to EDMs from complex flavour-violating couplings of the Higgs boson have received notable attention lately (see~e.g.~\cite{Harnik:2012pb,Goudelis:2011un,Blankenburg:2012ex}), while, to the best of our knowledge, the bounds on the $tuh$ couplings (\ref{eq:L}) that arise from the EDM of the neutron have only been considered in \cite{Harnik:2012pb}. Our work refines this analysis and extends it to the case of the $tch$ interactions. In both cases we resum large leading logarithms, which in the latter case requires to perform a two-loop calculation, while considering one-loop effects is sufficient in the former case. We also present a systematic study of direct as well as indirect CP violation in the $D$-meson sector that is induced by the FCNC Higgs-boson couplings (\ref{eq:L}). These calculations allow us to derive model-independent bounds on certain combinations of the flavour-changing couplings $Y_{tq}$ and $Y_{qt}$ that apply to all BSM scenarios where the observables under consideration receive the dominant contribution from FCNC  interactions involving the Higgs boson and the top quark.
 
The outline of this article is as follows. In Section \ref{sec:modelindependent} we deduce model-independent constraints on the $tch$ and $tuh$ couplings that arise from direct and indirect probes. Our conclusions are presented in Section~\ref{sec:conclusions}. In Appendix~\ref{app:ew} we estimate the size of  electroweak corrections to hadronic EDMs, while in Appendix~\ref{app:other} indirect  bounds arising from FCNCs involving down-type quarks are studied. Matching corrections to the Weinberg operator related to the exchange of a neutral and a charged scalar are presented in Appendix~\ref{app:weinberg}.
  
 \section{\boldmath Model-independent analysis}
\label{sec:modelindependent}

Below we will derive model-independent bounds on the  $t c h$ and $tu h$ couplings. In Section~\ref{sec:direct} will review the existing limits that are provided by the current LHC data. The expected future sensitivity that may arise from a high-luminosity upgrade of the LHC~(HL-LHC) is also discussed.  The calculations needed to derive the indirect constraints that result from the non-observation of hadronic EDMs are presented in Sections~\ref{sec:indirecttch} and~\ref{sec:indirecttuh}. The limits that stem from $D$-meson physics are examined in Section~\ref{sec:up}. Like in the case of the collider bounds we will also discuss  the future prospects of the indirect constraints. 

\subsection{\boldmath LHC  constraints}
\label{sec:direct}

In the presence of~(\ref{eq:L}) and assuming that the branching ratio of $t \to bW$ is close to unity, one obtains for the $t \to c(u)h$    branching fractions (see~e.g.~\cite{Casagrande:2008hr})
\begin{eqnarray} \label{eq:Brtch}
\begin{split}
{\rm Br} \left ( t \to q h \right ) & \simeq \frac{\sqrt{2}}{4 G_F \hspace{0.25mm} m_W^2} \frac{(1-x_{h/t})^2 \hspace{0.25mm} x_{W/t}}{(1-x_{W/t})^2 \hspace{0.25mm} (1+2x_{W/t})} \! \left [ |Y_{tq}|^2 + |Y_{qt}|^2 + \frac{4 \sqrt{x_{q/t}}}{1-x_{h/t}} \, {\rm Re} \left (Y_{tq} Y_{qt} \right ) \right ]  \hspace{5.5mm} \\[3mm]
& \simeq 0.26 \, \Big [ |Y_{tq}|^2 + |Y_{qt}|^2  \Big] \,,
\end{split}
\end{eqnarray}
with $q = c, u$. Here $x_{a/b} \equiv m_a^2/m_b^2$ and we have employed $G_F = 1.167 \cdot 10^{-5} \, {\rm GeV}^{-2}$ as well as the pole masses $m_t = 173.2 \, {\rm GeV}$, $m_h = 125 \, {\rm GeV}$ and $m_W = 80.4 \, {\rm GeV}$ to obtain the final result. Furthermore, the numerically subleading term proportional to ${\rm Re} \left (Y_{tq} Y_{qt} \right )$ in the first line of (\ref{eq:Brtch}) has been neglected.

Recently the CMS collaboration performed a search for $t \to qh$ which is based on $19.5 \, {\rm fb}^{-1}$ of $\sqrt{s} = 8 \, {\rm TeV}$ data and uses a combination of multilepton  and diphoton plus lepton final states  \cite{CMSPASHIG-13-034}. These measurements result in the following 95\% confidence level~(CL) upper limits 
\beq \label{eq:bestBRpresent}
{\rm Br} \left ( t \to c h \right )  < 0.56\% \,, \qquad  {\rm Br} \left ( t \to u h \right )  < 0.45\% \,.
\eeq 
where the former bound has been derived in \cite{CMSPASHIG-13-034}, while the latter exclusion has been found in \cite{Greljo:2014dka}. The quoted bounds translate into the limits 
\beq \label{eq:sqrtYYlimitnow}
\sqrt{|Y_{tc}|^2 + |Y_{ct}|^2} \lesssim 0.14 \,, \qquad \sqrt{|Y_{tu}|^2 + |Y_{ut}|^2} \lesssim 0.13 \,,
\eeq
on the $tch$ and $tuh$  couplings entering (\ref{eq:L}). Bounds complementary to those given in~(\ref{eq:bestBRpresent})  have been obtained by the ATLAS collaboration \cite{Aad:2014dya} and in \cite{Craig:2012vj}. Using a data sample corresponding to an integrated luminosity of $20.3 \, {\rm fb}^{-1}$ at $\sqrt{s} = 8 \, {\rm TeV}$ and $4.7 \, {\rm fb}^{-1}$ at $\sqrt{s} = 7 \, {\rm TeV}$, ATLAS utilised the $h \to \gamma \gamma$ channel to arrive at the 95\% CL bound ${\rm Br} \left ( t \to c h \right )  < 0.83\%$, which implies $\sqrt{|Y_{tc}|^2 + |Y_{ct}|^2} \lesssim 0.18$. The analysis \cite{Craig:2012vj} finally  infers a limit  ${\rm Br} \left ( t \to c h \right )  < 2.7\%$ at 95\% CL by performing a recast of a CMS anomalous multilepton search which utilises $4.7 \, {\rm fb}^{-1}$ of $\sqrt{s} = 7 \, {\rm TeV}$ data \cite{Chatrchyan:2012mea}. The corresponding limit on the FCNC top-quark couplings reads $\sqrt{|Y_{tc}|^2 + |Y_{ct}|^2} \lesssim 0.32$. 

Constraints on the interactions in (\ref{eq:L}) also derive from vector boson plus Higgs production as recently emphasised in \cite{Greljo:2014dka}. Employing the CMS data samples data corresponding to integrated luminosities of $19.5 \, {\rm fb}^{-1}$ at $\sqrt{s} = 8 \, {\rm TeV}$ and $5.0 \, {\rm fb}^{-1}$ at $\sqrt{s} = 7 \, {\rm TeV}$ \cite{CMS-PAS-HIG-13-053}, one obtains ${\rm Br} \left (t \to ch \right) \lesssim 1.2\%$ and ${\rm Br} \left (t \to uh \right) \lesssim 0.7\%$ corresponding to $\sqrt{|Y_{tc}|^2 + |Y_{ct}|^2} \lesssim 0.21$ and $\sqrt{|Y_{tu}|^2 + |Y_{ut}|^2} \lesssim 0.16$ at 95\% CL   \cite{Greljo:2014dka}. Additional information on the FCNC top-Higgs couplings can also be gained from the non-observation  of anomalous single top-quark production (see~e.g.~\cite{Harnik:2012pb}). Recasting the CMS multilepton search  which employs $19.5 \, {\rm fb}^{-1}$ of data collected at $\sqrt{s} = 8 \, {\rm TeV}$ \cite{CMSPASSUS-13-002}, the analysis \cite{Greljo:2014dka} finds the 95\% CL upper limits ${\rm Br} \left (t \to ch \right) \lesssim 1.5\%$ and ${\rm Br} \left (t \to uh \right) \lesssim 1.0\%$, which give rise to the bounds $\sqrt{|Y_{tc}|^2 + |Y_{ct}|^2} \lesssim 0.23$ and $\sqrt{|Y_{tu}|^2 + |Y_{ut}|^2} \lesssim 0.19$. Notice that the latter limits  are competitive with the one found in (\ref{eq:sqrtYYlimitnow}) from $t \to jh$ decays. 

At a HL-LHC with an integrated luminosity of $3000 \, {\rm fb}^{-1}$ at $\sqrt{s} = 14 \, {\rm TeV}$, the bounds on flavour-changing top-quark decays are expected to improve significantly. For instance, the study \cite{Agashe:2013hma} quotes the  following 95\% CL upper limit
\beq \label{eq:HLHC}
{\rm Br} \left (t \to  q h \right) \lesssim 2.0 \cdot 10^{-4} \,,
\eeq
based on the multilepton sample, i.e.~$t \bar t \to bW+ qh \to b\ell\nu+q\ell\ell X$. A slightly weaker bound of ${\rm Br} \left (t \to qh \right) \lesssim 5 \cdot 10^{-4}$ at 95\% CL  \cite{Agashe:2013hma} seems to be attainable in the diphoton channel $t \bar t \to bW+ qh \to b\ell\nu+q\gamma\gamma$ (see also \cite{ATL-PHYS-PUB-2013-012} for a recent study). Using (\ref{eq:Brtch}) the HL-LHC projection~(\ref{eq:HLHC}) leads to 
\beq \label{eq:sqrtYYlimitfuture}
\sqrt{|Y_{tq}|^2 + |Y_{qt}|^2} \lesssim 2.8 \cdot 10^{-2} \,. 
\eeq
This limit sets the future standard that any other constraint on the FCNC couplings~(\ref{eq:L}) has to be compared to.  Further recent collider studies of flavour-changing Higgs-boson couplings to the top quark can be found in \cite{Wang:2012gp,Chen:2013qta,Atwood:2013ica,Zhang:2013xya}.

\subsection{\boldmath EDM constraints  on top-charm-Higgs couplings}
\label{sec:indirecttch}

Integrating out the top quark and the Higgs boson at a scale $\mu_t = {\cal O} (m_t)$, the FCNC couplings~(\ref{eq:L}) lead to  effective interactions of the form 
\begin{equation} \label{eq:Leff}
\begin{split}
{\cal L}_{\rm eff} \supset & - d_q (\mu_t)\, \frac{i}{2} \, \bar q  \hspace{0.25mm}  \sigma^{\mu\nu} \gamma_5 \hspace{0.25mm} q \, F_{\mu\nu} - \tilde d_q (\mu_t)\, \frac{ig_s (\mu_t)}{2} \, \bar q  \hspace{0.25mm}   \sigma^{\mu\nu}  T^a \gamma_5 \hspace{0.25mm} q \, G_{\mu\nu}^a \\[2mm] & - w (\mu_t)\, \frac{1}{3} \hspace{0.25mm} f^{abc} \, G_{\mu \sigma}^a 
G_{\nu}^{b, \sigma} \widetilde G^{c, \mu \nu} \,,
\end{split}
\end{equation}
where $q=u,d$, while $\widetilde G^{a,\mu\nu}=\frac12\epsilon^{\mu\nu\alpha\beta}\,G_{\alpha\beta}^a$ is the dual field-strength tensor of QCD, with $\epsilon^{\mu \nu \lambda \rho}$ the fully anti-symmetric Levi-Civita tensor ($\epsilon^{0123} = 1$). $T^a$ are the colour generators normalised as ${\rm Tr} \hspace{0.5mm} (T^a T^b) = \delta^{ab}/2$. 

\begin{figure}[!t]
\begin{center}
\vspace{-5mm}
\includegraphics[height=0.25 \textwidth]{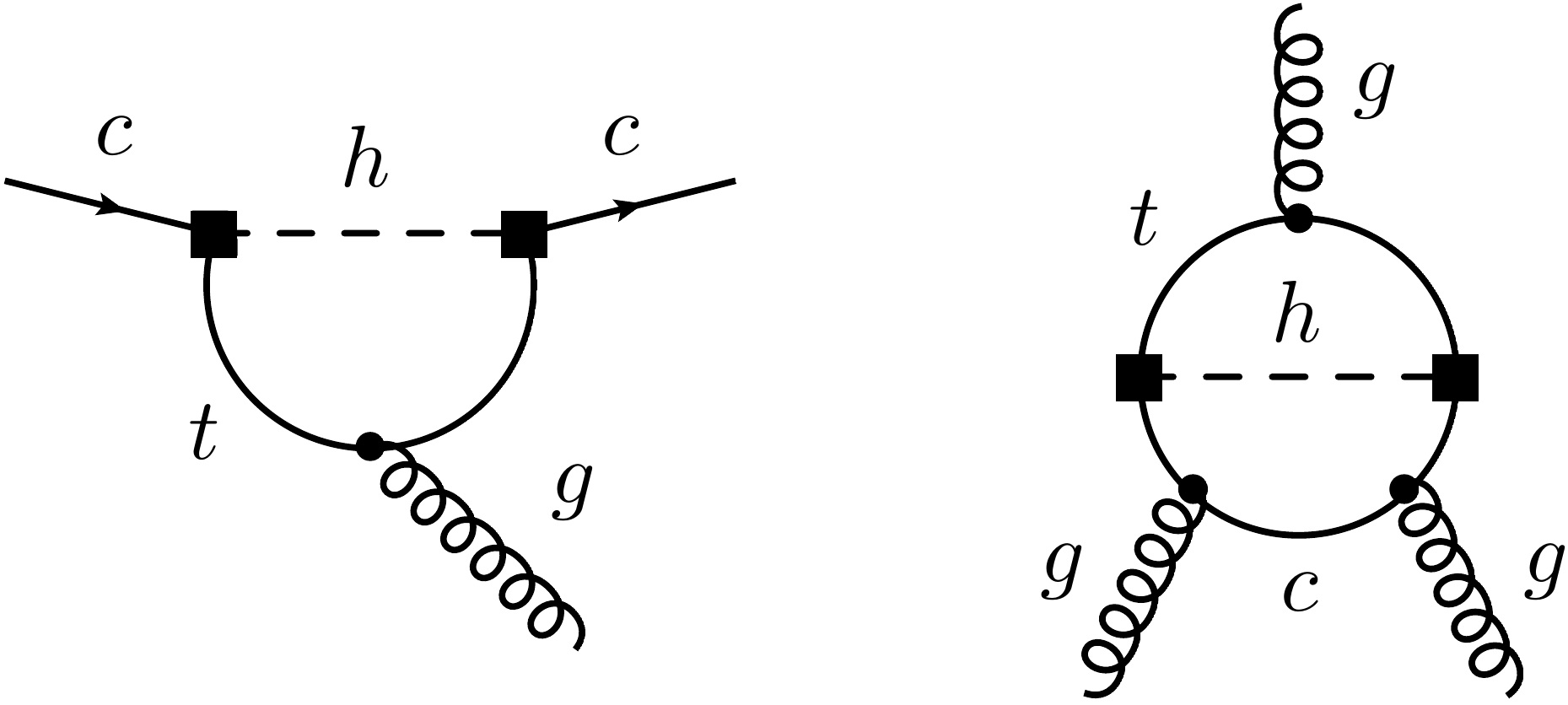}
\vspace{2mm}
\caption{\label{fig1} Left: One-loop  contribution to the CEDM of the charm quark.  Right: Example of a two-loop contribution to the Weinberg operator. The black squares in both graphs indicate the $tch$ couplings introduced in (\ref{eq:L}).}
\end{center}
\end{figure}

The initial condition  $\tilde d_c(\mu_t)$ of the charm-quark chromoelectric dipole moment (CEDM) is obtained from the one-loop diagram shown on the left-hand side of~Figure~\ref{fig1}. A straightforward calculation gives 
\beq \label{eq:CEDMc}
\tilde d_c(\mu_t)= \frac{1}{32 \pi^2} \frac{m_t}{m_h^2} \, f_1 (x_{t/h}) \, {\rm Im} \left ( Y_{tc} Y_{ct} \right ) \,, 
\eeq
with 
\beq \label{eq:f}
f_1 (x) = \frac{x-3}{(x-1)^2} + \frac{2}{(x-1)^3} \hspace{0.25mm} \ln x \,.
\eeq
Our result  (\ref{eq:CEDMc}) agrees with the findings of \cite{Boyd:1990,Jung:2013hka}. It furthermore resembles the expressions given in~\cite{Harnik:2012pb,Blankenburg:2012ex} after a suitable replacement of charge factors and couplings as well as taking the limit $x_{t/h} \to 0$. Notice that $\tilde d_c(\mu_t)$ is enhanced by the top-quark mass which provides the necessary chirality flip to generate a dipole transition. 

Finding the matching condition $w(\mu_t)$ of the Weinberg operator requires the computation of two-loop diagrams in the full theory. An example graph is display on the right in~Figure~\ref{fig1}. Setting the charm-quark mass  to zero, we obtain the compact expression 
\beq \label{eq:weinberg}
w (\mu_t)=  \frac{g_s^3 (\mu_t)}{(32 \pi^2)^2} \, \frac{1}{m_h^2} \, f_2 (x_{t/h}) \, {\rm Im} \left ( Y_{tc} Y_{ct} \right )  \,,
\eeq
with 
\beq \label{eq:newg}
f_2(x) = -\frac{x^2-5 x-2}{3 \left (x-1 \right )^3}-\frac{2x}{ \left (x-1 \right )^4} \hspace{0.25mm}  \ln x \,.
\eeq
To cross-check our result (\ref{eq:weinberg}) we have also calculated the two-loop contribution to the neutron EDM from neutral and charged Higgs-boson exchange  finding perfect agreement with the original computations~\cite{Weinberg:1989dx,Dicus:1989va} (see  also  \cite{Jung:2013hka} for a recent  discussion). For completeness we present the corresponding analytic results in Appendix~\ref{app:weinberg}.

\begin{figure}[!t]
\begin{center}
\vspace{-5mm}
\includegraphics[height=0.25 \textwidth]{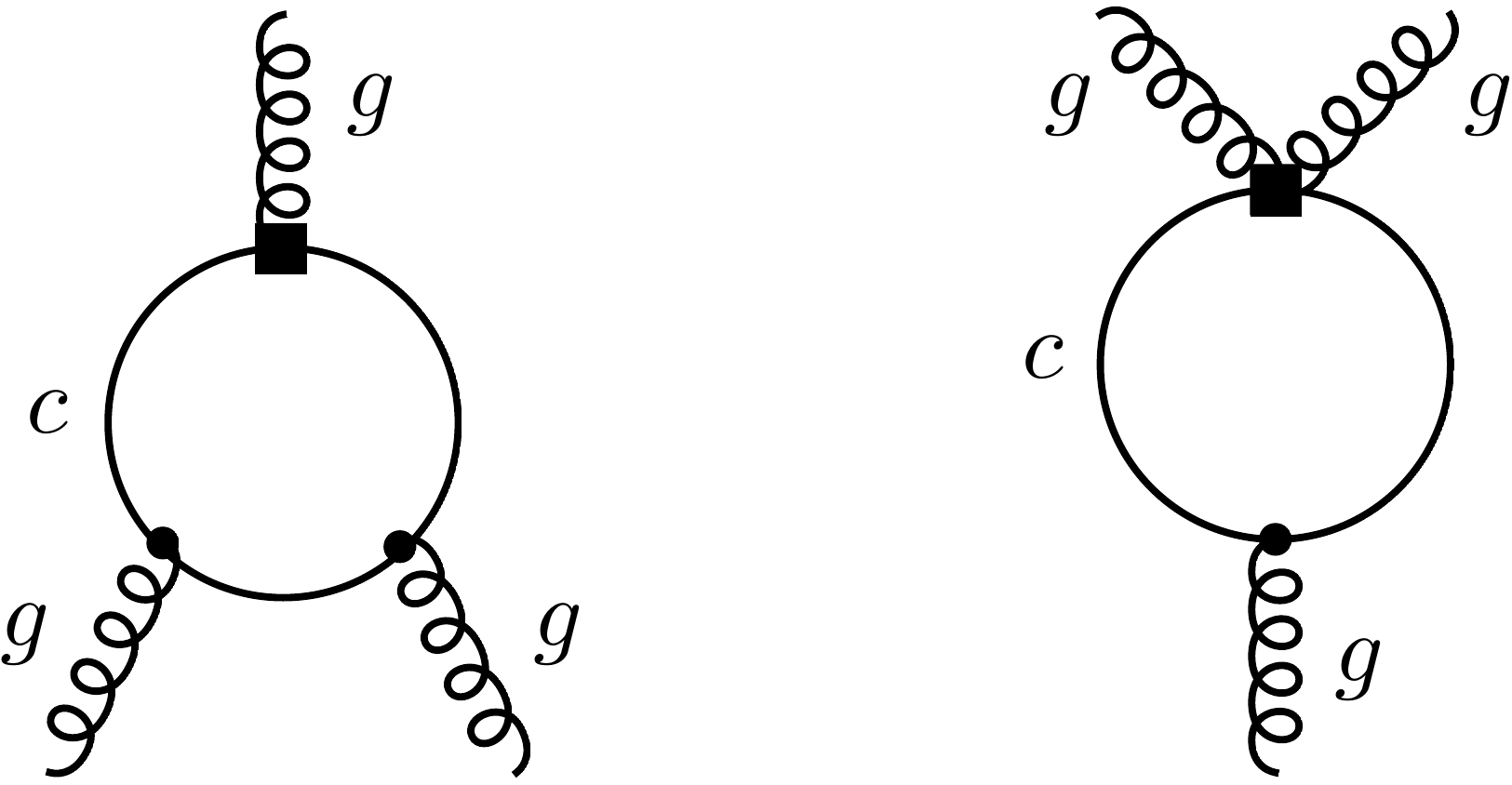}
\vspace{2mm}
\caption{\label{fig2} One-loop diagrams leading to a correction to the Weinberg operator at the charm-quark threshold. The black square denotes the insertion of the charm-quark CEDM.}
\end{center}
\end{figure}

The Weinberg operator in (\ref{eq:Leff}) mixes under renormalisation into the quark EDMs and CEDMs, while the opposite is not the case. However, the coefficient $w$  receives a finite  matching correction at each  heavy-quark threshold from the corresponding CEDM.   At the one-loop level, one finds  \cite{Boyd:1990,Braaten:1990gq,Chang:1990jv}
\beq \label{eq:threshold}
\delta w (m_c) = \frac{g_s^3 (m_c)}{32 \pi^2} \, \frac{\tilde d_c (m_c)}{m_c} \,,
\eeq
when the charm quark is integrated out. Here $m_c = 1.3 \, {\rm GeV}$ is the $\overline{\rm MS}$ charm-quark mass. The relevant diagrams are given in Figure~\ref{fig2}. The phenomenological importance of the charm-quark threshold correction (\ref{eq:threshold}) has  been stressed recently in \cite{Sala:2013osa} (see also~\cite{Jung:2013hka,Kamenik:2011dk,Brod:2013cka} for related discussions of the relevance of the top-quark and bottom-quark threshold corrections). Notice that the initial condition $w (\mu_t)$ is compared to the  threshold correction $\delta w(m_c)$ suppressed by a factor of $m_c/m_t \simeq 1/125$. As we will see below this implies that the contributions to the hadronic EDMs arising from the Weinberg operator are fully  dominated by infrared physics associated to scales of the order of the charm-quark mass. 

The combined effects of the finite shift (\ref{eq:threshold}) in $w (m_c)$ and the subsequent renormalisation group~(RG) evolution (see~e.g.~\cite{Degrassi:2005zd}) to the hadronic scale $\mu_H = 1 \, {\rm GeV}$ will induce non-zero contributions also for the EDMs and CEDMs of the down quark and up quark. Performing 5-flavour, 4-flavour and 3-flavour running, we obtain in leading-logarithmic (LL) approximation 
\beq \label{eq:charmcoefficients}
\begin{split}
\frac{d_d (\mu_H)}{e} & = 2.3 \cdot 10^{-8} \, \tilde d_c(m_t)+1.0 \cdot 10^{-4} \, {\rm GeV} \, w(m_t)\,, \\[1mm]
\frac{d_u (\mu_H)}{e} & = -2.1 \cdot 10^{-8} \, \tilde d_c(m_t)-9.1 \cdot 10^{-5} \, {\rm GeV} \, w(m_t)\,, \\[2mm]
\tilde d_d (\mu_H) & = 1.8 \cdot 10^{-6} \,  \tilde d_c(m_t)+ 7.0  \cdot 10^{-4} \, {\rm GeV} \, w(m_t)\,, \\[3mm]
\tilde d_u (\mu_H) & = 8.2 \cdot 10^{-7} \,  \tilde d_c(m_t)+ 3.1  \cdot 10^{-4} \, {\rm GeV} \, w(m_t)\,, \\[3mm]
w(\mu_H) & = 1.7 \cdot 10^{-2} \, {\rm GeV}^{-1} \,  \tilde d_c(m_t)+ 0.41 \, w(m_t)\,,
\end{split}
\eeq
where we have identified $\mu_t = m_t =163.3 \,{\rm GeV}$. The given numerical coefficients correspond to $\alpha_s (m_t)= 0.109$, $\alpha_s (m_b) = 0.213$, $\alpha_s (m_c) = 0.319$, $\alpha_s (\mu_H) = 0.362$, $m_d (\mu_H) = 5.4 \cdot 10^{-3} \, {\rm GeV}$ and $m_u (\mu_H) = 2.4 \cdot 10^{-3} \, {\rm GeV}$. We recall that the RG evolution tends to suppress  the coefficients $d_q (\mu)$, $\tilde d_q (\mu)$ and $w (\mu)$. Terminating the running at $\mu_H = 1 \, {\rm GeV}$ rather than at the charm-quark threshold $m_c = 1.3  \, {\rm GeV}$ hence leads to weaker constraints on~$\left | {\rm Im} \left ( Y_{tc} Y_{ct} \right )\right |$. The same is true in the case of the EDM constraints on the $tuh$ couplings that will be discussed in the next subsection. 

In terms of $d_q (\mu_H)$, $\tilde d_q (\mu_H)$ and $w (\mu_H)$  the neutron EDM \cite{Pospelov:2005pr} takes the following  form 
\begin{equation} \label{eq:dn}
\begin{split}
\frac{d_n}{e} & = (1.0 \pm 0.5) \left \{ 1.4\left  [\frac{d_d (\mu_H)}{e} - 0.25 \, \frac{d_u  (\mu_H)}{e} \right ] + 1.1 \left  [\tilde d_d (\mu_H) + 0.5 \, \tilde d_u  (\mu_H) \right ] \right \} \\[3mm] 
& \phantom{xx}+  (22 \pm 10) \cdot 10^{-3} \, {\rm GeV} \, w (\mu_H) \,, 
\end{split}
\eeq
while for deuteron \cite{Lebedev:2004va} one has
\begin{equation} \label{eq:dD}
\begin{split}
\frac{d_D}{e} & =(0.5 \pm 0.3)  \left [ \frac{d_d  (\mu_H)}{e} + \frac{d_u  (\mu_H)}{e} \right ]  + \Big [ 5^{+11}_{-3} + \left ( 0.6 \pm 0.3 \right ) \Big ]  \big (\tilde d_d (\mu_H) - \tilde d_u (\mu_H) \big ) \\[3mm] 
& \phantom{xx} - (0.2\pm 0.1) \hspace{0.25mm} \big (\tilde d_d (\mu_H) + \tilde d_u (\mu_H) \big ) +  (22 \pm 10) \cdot 10^{-3} \, {\rm GeV} \, w (\mu_H) \,.
\end{split}
\end{equation} 

Inserting the results (\ref{eq:charmcoefficients}) into the general expression (\ref{eq:dn}) for the  EDM of the neutron, we then find 
\beq \label{eq:dnoe}
\left | \frac{d_n}{e} \right |   =  \left | 2.0 \cdot 10^{-4} \, \tilde d_c(m_t)+ 5.4 \cdot 10^{-3} \, {\rm GeV} \, w (m_t)\right |  \simeq 5.8 \cdot 10^{-23} \, {\rm cm} \;  \left | {\rm Im} \left ( Y_{tc} Y_{ct} \right ) \right |  \,.
\eeq
In order to obtain conservative bounds on the $tch$ couplings, we have set the numerical coefficients $(1.0 \pm 0.5)$ and $(22 \pm 10) \cdot 10^{-3} \, {\rm GeV}$ in (\ref{eq:dn}) to $0.5$ and $12 \cdot 10^{-3} \, {\rm GeV}$. The  final result in (\ref{eq:dnoe})  also holds for $|d_D/e|$ which implies that in the case of a non-zero charm-quark CEDM the contribution to (\ref{eq:dD})  from the Weinberg operator is the by far largest correction. Notice furthermore that the contribution from the Weinberg operator itself is almost entirely due to the threshold correction $\delta w (m_c)$ with the initial condition~$w (m_t)$ amounting  to a relative effect of less than $1 \permil$ only. 

The present 90\% CL bound on the EDM of the neutron reads \cite{Baker:2006ts}
\beq \label{eq:dnoepresent}
\left | \frac{d_n}{e} \right |  < 2.9 \cdot 10^{-26} \, {\rm cm}  \,,
\eeq
while a limit on $|d_D/e|$ does not exist at the moment. Using (\ref{eq:dnoe}) the latter bound can be translated into the inequality 
\beq \label{eq:sqrtImYYpresent}
\left | {\rm Im} \left ( Y_{tc} Y_{ct} \right ) \right |  \lesssim 5.0 \cdot 10^{-4} \,.
\eeq

The expected future sensitivities on the EDM of the neutron is $|d_n/e| \lesssim 10^{-28} \, {\rm cm}$ \cite{Hewett:2012ns}, whereas in the case of deuteron even a limit of 
\beq \label{eq:dDoefuture}
\left | \frac{d_D}{e} \right | \lesssim 10^{-29} \, {\rm cm} \,,
\eeq
might be achievable  in the long run  \cite{dDfuture}. Such a precision would allow to set a bound of 
\beq  \label{eq:sqrtImYYfuture}
\left | {\rm Im} \left ( Y_{tc} Y_{ct} \right ) \right |  \lesssim 1.7 \cdot 10^{-7} \,.
\eeq
A factor $300$ improvement of the current bound (\ref{eq:dnoepresent}) on the neutron EDM  would instead result in  $\left | {\rm Im} \left ( Y_{tc} Y_{ct} \right ) \right | \lesssim 1.7 \cdot 10^{-6}$. 

\subsection{\boldmath EDM constraints  on top-up-Higgs couplings}
\label{sec:indirecttuh}

In the case of the $tuh$ couplings the calculation of the EDM of the neutron and deuteron is simplified by the fact that a up-quark EDM and CEDM is already generated from integrating out the top quark and the Higgs boson. The initial condition $d_u (\mu_t)$ of the up-quark EDM  is obtained from a one-loop diagram similar to the graph on the left-hand side of Figure~\ref{fig1}, but with the gluon replaced by a photon. In accordance with the literature~\cite{Harnik:2012pb,Blankenburg:2012ex,Boyd:1990,Jung:2013hka}, we find
\beq \label{eq:dumatch}
d_u (\mu_t)=  \frac{Q_u \hspace{0.25mm} e}{32 \pi^2} \frac{m_t}{m_h^2} \, f_1 (x_{t/h}) \, {\rm Im} \left ( Y_{tu} Y_{ut} \right ) \,.
\eeq
Here $Q_u = 2/3$ denotes the electric charge of the up quark and the loop function $f_1$ has been given in (\ref{eq:f}). The result for the matching corrections $\tilde d_u (\mu_t)$ and $w (\mu_t)$ are readily obtained from (\ref{eq:CEDMc}) and (\ref{eq:weinberg}) by the replacements $Y_{tc} \to Y_{tu}$ and $Y_{ct} \to Y_{ut}$ of the FCNC Higgs-boson couplings. Since  $w(\mu_t)$ has numerically a negligible impact on the hadronic EDMs, we set this contribution to zero in what follows.  

At LL accuracy the up-quark EDM and CEDM at the hadronic scale are given in terms of the high-scale coefficients by
\beq \label{eq:upcoefficients}
\begin{split}
\frac{d_u (\mu_H)}{e}  = 0.82 \; \frac{d_u (m_t)}{e} - 0.46 \, \tilde d_u (m_t)\,, \qquad \tilde d_u (\mu_H)  = 0.90 \, \tilde d_u (m_t)\,, 
\end{split}
\eeq
while $d_d(\mu_H) = \tilde d_d(\mu_H) = w (\mu_H) = 0$ at the one-loop order.  Inserting these results into~(\ref{eq:dn}), we arrive at 
\beq \label{eq:dnoeup}
\left | \frac{d_n}{e} \right |   =  \left | -0.14\, \frac{d_u(m_t)}{e} + 0.33 \, \tilde d_u (m_t)\right |   \simeq 6.7 \cdot 10^{-20} \, {\rm cm} \;  \left | {\rm Im} \left ( Y_{tu} Y_{ut} \right ) \right |  \,, 
\eeq
in the case of the neutron and at 
\beq \label{eq:dDoeup}
\left | \frac{d_D}{e} \right |   =  \left |  \hspace{0.25mm} 0.66 \; \frac{d_u(m_t)}{e} - 2.54  \, \tilde d_u (m_t)\right | \simeq 6.0 \cdot 10^{-19} \, {\rm cm} \;  \left | {\rm Im} \left ( Y_{tu} Y_{ut} \right ) \right |  \,, 
\eeq
for deuteron. In order to obtain these semi-analytic expressions we have set the  factor $(1.0 \pm 0.5)$ entering the expression for $d_n/e$ to $0.5$  and chose the multiplicative factors in~(\ref{eq:dD})  such that the cancellation between the up-quark EDM and CEDM in $d_D/e$ is maximised. Using the formulas (\ref{eq:dnoeup}) and (\ref{eq:dDoeup}) to bound  $\left |{\rm Im} \left ( Y_{tu} Y_{ut} \right ) \right |$  results therefore in conservative upper limits. 

It follows that the existing limit (\ref{eq:dnoepresent}) on the neutron EDM already give rise to the very strong limit
\beq \label{eq:upnow}
\left | {\rm Im} \left ( Y_{tu} Y_{ut} \right ) \right |  \lesssim 4.3 \cdot 10^{-7} \,.
\eeq
 This bound is weaker by a factor of around 10 than the limit quoted in \cite{Harnik:2012pb}. This discrepancy can be resolved by noticing that in the latter article only the contribution of $d_u$ to $d_n$ is included, large logarithms are not resummed and a naive estimate of the neutron EDM is utilised. All these approximations tend to enhance the impact of the $tuh$ couplings~(\ref{eq:L}) on the prediction of~$d_n$. 

By achieving the sensitivity (\ref{eq:dDoefuture}) on the EDM of deuteron this bound would further improve, leading to 
\beq \label{eq:upfuture}
\left | {\rm Im} \left ( Y_{tu} Y_{ut} \right ) \right |  \lesssim 1.7 \cdot 10^{-11} \,.
\eeq
Neutron EDM measurements, on the other hand, are expected to reach a sensitivity of $\left | {\rm Im} \left ( Y_{tu} Y_{ut} \right ) \right |  \lesssim 1.5 \cdot 10^{-9}$ within a time scale of a few years. 

\subsection{Constraints from  FCNC charm-up transitions}
\label{sec:up}

Certain combinations of the couplings (\ref{eq:L}) such as the product $Y_{ut}^\ast Y_{tc}$ can be probed by  $D$-meson physics. In what follows we will consider both CP violation in the $\Delta C =1$ and $\Delta C = 2$ sectors. 

In the former case a useful constraint arises from the difference $\Delta A_{\rm CP}$ between the two direct CP asymmetries in $D \to K^+ K^-$ and $D \to \pi^+ \pi^-$. This observable can receive sizeable corrections from the chromomagnetic dipole operator 
\beq \label{eq:Q8}
Q_8 = \frac{g_s}{(4 \pi)^2} \, m_c \, \bar u_L \sigma^{\mu \nu} T^a c_R \hspace{0.25mm} G_{\mu \nu}^a \,,
\eeq
and its chirality-flipped partner $\tilde Q_8$ obtained by $L \leftrightarrow R$. These operators appear in the effective $\Delta C =1$ Lagrangian as follows ${\cal L}_{\rm eff} \supset -4 G_F/\sqrt{2} \, ( C_8 \hspace{0.25mm} Q_8 + \tilde C_8 \hspace{0.25mm} \tilde Q_8)$.  The initial condition of the Wilson coefficient of $Q_8$ is determined from a one-loop diagram similar to the graph on the left-hand side of Figure \ref{fig1}, but with the outgoing charm quark replaced by an up quark. We find 
\beq \label{eq:DC8}
\Delta C_8 (\mu_t) = \frac{\sqrt{2}}{16 \hspace{0.25mm} G_F} \, \frac{m_t}{m_c} \, \frac{Y_{ut}^\ast Y_{tc}}{m_h^2} \, f_1(x_{t/h}) \,,
\eeq
with $f_1$ given in (\ref{eq:f}). The result for the new-physics contribution $\Delta \tilde C_8 (\mu_t)$ is obtained from the above expression  by the interchange $Y_{ut} \leftrightarrow Y_{tc}$.  

Since the charm quark is too heavy for chiral perturbation theory to be applicable and too light for heavy-quark effective theory to be trusted, precise theoretical predictions in $D$-meson decays are notoriously difficult. As a result the bounds derived below are  plagued by ${\cal O} (1)$ uncertainties, which should be clearly kept in mind.  Following \cite{Isidori:2011qw} we write 
\beq \label{eq:DACPtheory}
|\Delta A_{\rm CP}| \simeq  \, 0.94\,  | \hspace{0.125mm} {\rm Im} \hspace{0.25mm} \big ( \Delta C_8 (m_t) \big  )  |  \simeq  \frac{\left | {\rm Im} \left ( Y_{ut}^\ast Y_{tc} \right ) \right |}{4.0 \cdot 10^{-4}} \, \% \,,
\eeq
where for simplicity we only have incorporated the contribution from $Q_8$. In order to set an upper bound on $\left | {\rm Im} \left ( Y_{ut}^\ast Y_{tc} \right ) \right |$, we will require that 
\beq  \label{eq:DACPexp}
|\Delta A_{\rm CP}| \lesssim 1 \% \,.
\eeq
which in view of the present world average $\Delta A_{\rm CP} = -(0.33 \pm 0.12)\%$ \cite{Amhis:2012bh} seems like a conservative choice. It follows that  
\beq \label{eq:DACPbound}
 \left | {\rm Im} \left ( Y_{ut}^\ast Y_{tc} \right ) \right | \lesssim 4.0 \cdot 10^{-4} \,.
\eeq
Since the future sensitivity of  $|\Delta A_{\rm CP}|$ to the $tch$ and $tuh$ couplings is largely dependent on theoretical progress concerning the understanding of hadronic physics, which is hard to predict, we do not attempt to make any projection in this case.  

In the presence of (\ref{eq:L}) the $D$--$\bar D$ mixing amplitude receive contributions from box diagrams with Higgs-boson and top-quark exchange. The phenomenologically most important contribution comes from the mixed-chirality operator 
\beq \label{eq:Q4} 
Q_4 = (\bar c_L u_R) (\bar c_R u_L) \,, 
\eeq
due to its large anomalous dimension and a chiral enhancement of its hadronic matrix element. Normalising the associated effective Lagrangian as ${\cal L}_{\rm eff} \supset - 4 G_F/\sqrt{2} \, C_4 \hspace{0.25mm} Q_4$, we find  in agreement with \cite{Harnik:2012pb} the matching correction 
\beq \label{eq:DC4}
\Delta C_4 (\mu_t) =\frac{1}{32 \pi^2} \, \frac{\sqrt{2}}{4 G_F} \frac{f_3(x_{t/h})}{m_h^2}  \,  Y_{tc}^\ast Y_{ut}^\ast  Y_{tu} Y_{ct}   \,,
\eeq
with 
\beq \label{eq:g}
f_3(x) = \frac{4x}{(x-1)^2} - \frac{2x^2+2x}{(x-1)^3} \, \ln x \,.
\eeq

Allowing the new-physics  contribution to saturate the experimental bounds on CP violation in $D$--$\bar D$ mixing, one arrives at \cite{Gedalia:2009kh}
\beq \label{eq:CPinD}
|{\rm Im} \left ( \Delta C_4 (m_t) \right ) | \lesssim 3.3 \cdot 10^{-10} \,,
\eeq
which corresponds to a bound of 
\beq \label{eq:boundDmix}
\sqrt{\left | {\rm Im} \left (  Y_{tc}^\ast Y_{ut}^\ast   Y_{tu} Y_{ct}   \right ) \right | } \lesssim 4.1 \cdot 10^{-4} \,.
\eeq
Note that in \cite{Harnik:2012pb} a bound on $\sqrt{\left |  Y_{tc}^\ast Y_{ut}^\ast   Y_{tu} Y_{ct} \right | }$ has been derived which is a factor of about $\sqrt{5}$ weaker than the latter limit. This is a simple consequence of the fact (see~e.g.~\cite{Gedalia:2009kh}) that the constraint arising from CP violation in $D$-meson mixing is by a factor of 5 stronger than the bound coming from the CP-conserving measurements. 

Future measurements of CP violation in $D$--$\bar D$ mixing at LHCb \cite{Bediaga:2012py} and Belle~II~\cite{Aushev:2010bq} are expected to improve the current bound (\ref{eq:CPinD}) by at least a factor of 10.  Such an improvement would result in 
\beq \label{eq:boundDmixfuture}
\sqrt{\left | {\rm Im} \left (  Y_{tc}^\ast Y_{ut}^\ast   Y_{tu} Y_{ct}   \right ) \right | } \lesssim 1.3 \cdot 10^{-4} \,, 
\eeq
if one again allows the new-physics contribution (\ref{eq:DC4}) to saturated the future limit on CP violation in the $\Delta C =2$ sector. 

\subsection{\boldmath Summary of constraints}
\label{sec:summary}

\begin{table}[t!]
  \centering
  \parbox{13cm}{
  \begin{tabular}{@{\qquad}lccc@{\qquad}}
    Observable     & Coupling              &  Present bound         &  Future sensitivity      \\[1mm] \hline  \hline \\[-3mm]
         \multirow{2}{*}{LHC searches}
                  & $\sqrt{|Y_{tc}|^2 + |Y_{ct}|^2}$       &  $0.14$ &  $2.8 \cdot 10^{-2\phantom{0}}$ \\[1mm]
                   & $\sqrt{|Y_{tu}|^2 + |Y_{ut}|^2}$       &  $0.13$ &  $2.8 \cdot 10^{-2\phantom{0}}$ 
                  \\[2mm] \hline       \\[-3mm]            
     \multirow{2}{*}{$d_n$}
                  & $\left | {\rm Im} \left ( Y_{tc} Y_{ct} \right ) \right |$ &  $5.0 \cdot 10^{-4}$ & $1.7 \cdot 10^{-6\phantom{0}}$  \\
                  & $\left | {\rm Im} \left ( Y_{tu} Y_{ut} \right ) \right |$ &  $4.3 \cdot 10^{-7}$ & $1.5 \cdot 10^{-9\phantom{0}}$ \\ [1mm]            
      \multirow{2}{*}{$d_D$}
                  & $\left | {\rm Im} \left ( Y_{tc} Y_{ct} \right ) \right |$ &  --- & $1.7 \cdot 10^{-7\phantom{0}}$  \\
                  & $\left | {\rm Im} \left ( Y_{tu} Y_{ut} \right ) \right |$ & --- & $1.7 \cdot 10^{-11}$ 
                  \\[1mm]
                  \hline  \\[-3mm]            
     \multirow{1}{*}{$\Delta A_{\rm CP}$}
                  & $\left | {\rm Im} \left ( Y_{ut}^\ast Y_{ct} \right ) \right |$ &  $4.0 \cdot 10^{-4}$ & ---  \\ [1mm]            
      \multirow{1}{*}{$D$--$\bar D$ mixing}
                  & $\sqrt{\left | {\rm Im} \left (  Y_{tc}^\ast Y_{ut}^\ast   Y_{tu} Y_{ct}   \right ) \right | }$ & $4.1 \cdot 10^{-4}$ & $1.3 \cdot 10^{-4}$                    \\[2mm]
                  \hline                  
  \end{tabular}}
  \vspace{4mm}
  \caption{Summary of the most powerful constraints on the $tqh$ couplings with $q=c,u$. To obtain the 95\% CL upper limits we have assumed a Higgs-boson mass $m_h = 125 \, {\rm GeV}$ and neglected other possible contributions to the processes under considerations beyond those arising from (\ref{eq:L}). }
  \label{tab1}
\end{table}

In Table~\ref{tab1} we summarise the most stringent limits on the FCNC Higgs-boson couplings~(\ref{eq:L}) arising from collider physics (see Section~\ref{sec:direct}), hadronic EDMs   (see~Sections~\ref{sec:indirecttch} and \ref{sec:indirecttuh}) and CP violation in $D$-meson physics (see~Section~\ref{sec:up}). Whenever possible we give both the present bound and a projection of the future sensitivity. 

\section{\boldmath Conclusions}
\label{sec:conclusions}

The LHC discovery of the Higgs boson furnishes new opportunities in the search for physics beyond the SM. Since in the SM flavour-changing Higgs couplings to fermions are highly suppressed, discovering any evidence of a decay like  $t \to c h$ would strongly suggest the existence of new physics not  far above the TeV scale. In fact, both ATLAS and CMS have already provided their first limits on the $t \to c (u) h$ branching ratios (see~e.g.~\cite{CMSPASSUS-13-002,CMSPASHIG-13-034,Aad:2014dya,Craig:2012vj,Greljo:2014dka}). While these recent results still allow for branching ratios in excess of around 0.5\%, the searches for flavour-changing top-Higgs interactions will mature at the $14 \, {\rm TeV}$ LHC and it is expected that the current limits on the $t \to c (u) h$ branching ratios can be improved by roughly two order of magnitude. By achieving such a precision these searches would become sensitive to the~(maximal) $t \to ch$ rates predicted in an assortment of new-physics scenarios \cite{AguilarSaavedra:2004wm, Bejar:2000ub, Guasch:1999jp,Eilam:2001dh, Azatov:2009na,Casagrande:2010si, Agashe:2009di}. 

In this article we have emphasised the complementary between high-$p_T$ and low-energy precision measurements in extracting information about the properties of the flavour-changing top-Higgs  couplings. By considering a model-independent parameterisation of these interactions, we have obtained bounds on certain combinations of the $tch$ and  $tuh$ couplings that derive from the measurements of hadronic EDMs and CP-violating observables in the $D$-meson sector. While the limits  on the $tuh$ interactions due to the neutron EDM and charm-quark physics have been previously considered \cite{Harnik:2012pb},  our constraints on the $tch$ couplings are novel. The derivation of the latter bounds is based on a complete two-loop matching calculation and includes the resummation of large leading QCD logarithms by means of renormalisation group techniques. Given the model-independent character of our calculations, the derived limits can be used to constrain the parameter space of all beyond the SM scenarios where the considered quantities receive the dominant CP-violating contributions from flavour-changing  top-Higgs interactions. The presented results hence should prove useful in  monitoring the impact that further improved precision measurements of low-energy observable have in extracting information on the $t c h$  and $t u h$ couplings. 

\acknowledgments We are grateful to Filippo~Sala and Jure~Zupan for clarifying discussions concerning their works \cite{Sala:2013osa} and \cite{Harnik:2012pb}, respectively. MG acknowledges partial support by the UK Science~\& Technology Facilities Council~(STFC) under grant No.~ST/G00062X/1. UH would like thank the organisers of  ``The top-charm frontier at the LHC'' for an entertaining workshop that renewed his interest in flavour-changing top-Higgs interactions. He also acknowledges the warm hospitality and support of the CERN theory division.

\appendix

\section{Electroweak corrections to hadronic EDMs}
\label{app:ew}

In the presence of the $tch$ couplings (\ref{eq:L}) the EDM of the neutron and deuteron receive electroweak corrections at the two-loop level. The size of the induced effects can be estimated by inserting the effective charm-quark photon (gluon) interactions corresponding to $d_c$ ($\tilde d_c$) into the two one-loop graphs in which the photon (gluon) is emitted from the internal charm-quark line, i.e.~those with $W$-boson and would-be Goldstone boson exchange. 
Such a calculation leads to \cite{CorderoCid:2007uc}
\beq \label{eq:dd}
d_d  (\mu_H) \simeq -\frac{\alpha}{4 \pi} \frac{|V_{cd}|^2}{s_W^2} \frac{m_d \hspace{0.5mm} m_c}{m_W^2} \, d_c (\mu_t) \ln x_{c/W} \,,
\eeq
in LL approximation. An analogue expression holds for $\tilde d_d (\mu_H)$. Here $|V_{cd}| \simeq 0.22$ denotes the relevant CKM matrix element and $s_W^2 \simeq 0.23$ is the sine of the weak mixing angle.  Notice that $d_d$ is chirally suppressed by both the down-quark and charm-quark mass which signals that (\ref{eq:dd}) is formally a dimension-8 contribution. 

Numerically, one finds from (\ref{eq:dn}) that 
\beq \label{eq:dneEW}
\left | \frac{d_n}{e} \right |    \simeq 3.2 \cdot 10^{-28} \, {\rm cm} \;  \left | {\rm Im} \left ( Y_{tc} Y_{ct} \right ) \right |  \,,
\eeq
where for simplicity we have used $\alpha = 1/137$, evaluated the quark masses $m_d$ and $m_c$ at the hadronic scale and left the logarithm in (\ref{eq:dd}) unresummed. Comparing (\ref{eq:dneEW}) to (\ref{eq:dnoe}) we see that electroweak contributions to $d_n/e$ can be ignored for all practical purposes, because they are by more than five orders of magnitude smaller than the QCD effects. The same statement also holds in the case of the EDM of deuteron. These findings agree with those of \cite{Sala:2013osa}.  

\section{FCNC transitions in the down-type quark sector}
\label{app:other}

The $tch$ couplings in (\ref{eq:L}) can also be probed by quark FCNC transitions in the down-type quark sector.  Although the resulting constraints turn out to be not very restrictive, we will for completeness discuss as an example  the inclusive $B \to X_s \gamma$ decay.

In the case of $B \to X_s \gamma$ one has to consider both the EDM and CEDM interactions in~(\ref{eq:Leff}) as well as the magnetic dipole moment and chromomagnetic dipole moment  of the charm quark: 
\begin{equation} \label{eq:Leffmagnetic}
\begin{split}
{\cal L}_{\rm eff} \supset & - \mu_c (\mu_t) \, \frac{1}{2} \, \bar c  \hspace{0.25mm}  \sigma^{\mu\nu} c \, F_{\mu\nu} - \tilde \mu_c (\mu_t) \, \frac{g_s (\mu_t)}{2} \, \bar c  \hspace{0.25mm}   \sigma^{\mu\nu}  T^a   c \, G_{\mu\nu}^a \,.
\end{split}
\end{equation}

By employing the results of \cite{Hewett:1993em} we find that the new-physics contribution to  the Wilson coefficient of the electromagnetic dipole operator 
\beq \label{eq:Q7}
Q_7 = \frac{e}{(4 \pi)^2} \, m_b \, \bar s_L \sigma^{\mu \nu} b_R \hspace{0.25mm} F_{\mu \nu} \,,
\eeq
takes the form  
\beq \label{eq:DC7}
\Delta C_7 (\mu_t)  \simeq  \frac{m_c}{2 e} \, \frac{V_{cs}^\ast V_{cb}}{V_{ts}^\ast V_{tb}} \, \Big [  \mu_c (\mu_t) -  i \hspace{0.25mm} d_c  (\mu_t) \, \big ( 4  \ln x_{c/W} + 5 \big )  \Big  ]  \,.
\eeq
The expression for  $\Delta C_8 (\mu_t)$ which multiplies the chromomagnetic dipole operator $Q_8$ is obtained from (\ref{eq:DC7}) by the replacements $e \to 1$, $\mu_c (\mu_t) \to \tilde \mu_c (\mu_t)$ and $d_c (\mu_t) \to \tilde d_c (\mu_t)$. We see that to LL accuracy only the charm-quark EDM and CEDM contribute to $\Delta C_7 (\mu_t)$  and $\Delta C_8 (\mu_t)$, respectively. Numerically, the enhancement of the contribution of $d_c (\mu_t)$ $\big ( \tilde d_c (\mu_t) \big )$ with respect  to $\mu_c (\mu_t)$ $\big ( \tilde \mu_c (\mu_t) \big )$ amounts to a factor of around 28.   In our numerical analysis we therefore include only the LL terms. Since $\Delta C_8 (\mu_t)$ enters the predictions for the branching ratio of $B \to X_s \gamma$ first at the next-to-leading logarithmic order, we  neglect this contribution. Finally, we also identify $V_{cs}^\ast V_{cb} = - V_{ts}^\ast V_{tb}$, which holds to excellent approximation. 

Using now (\ref{eq:dumatch}), which also applies in the case of the charm quark, we obtain from~(\ref{eq:DC7}) in LL accuracy  
\beq \label{eq:C7dnoe}
\big |\Delta C_7 (m_t) \big | \simeq 2.1 \cdot 10^{-4} \, \left | {\rm Im} \left ( Y_{tc} Y_{ct} \right ) \right |  \lesssim 1.0 \cdot 10^{-7} \,,  
\eeq  
where in order to arrive at the final result we have utilised the current  bound (\ref{eq:dnoepresent}) on the neutron EDM. This result should be compared to the present 90\% CL limit 
\beq \label{eq:C7}
\big |\Delta C_7 (m_t) \big | < 0.37 \,,
\eeq
following  from a global analysis of $b \to s \gamma, \ell^+ \ell^-$ data \cite{Bobeth:2011st}.  Our bound is consistent with the result on the top-quark EDM $|d_t/e|$ derived in \cite{Kamenik:2011dk}, but weaker by a factor of around $m_t/m_c \simeq 125$ than the limit on $|d_c/e|$ quoted in \cite{Sala:2013osa}. The estimate (\ref{eq:C7dnoe}) shows clearly that indirect probes of the $tch$ couplings via quark FCNC transitions in the down-type quark sector will never be able to compete with the constraints arising from hadronic EDMs. 

\section{Two-loop matching corrections for the Weinberg operator}
\label{app:weinberg}

In this appendix we present the results for the matching corrections to the Weinberg operator resulting from Feynman diagrams involving the exchange of a neutral and a charged scalar. The expressions for the corresponding initial conditions have been calculated originally in the classic papers \cite{Weinberg:1989dx, Dicus:1989va}  in terms of twofold Feynman parameter integrals. Below we will give analytic results for these integrals. 

In the case of neutral scalar $S^0$ exchange, we parameterise the relevant interactions in the following way 
\beq
{\cal L} \supset - Y_{qq} \hspace{0.5mm} \bar q_L q_R  \hspace{0.5mm} S^0 + {\rm h.c.}  \,.
\eeq
Performing the matching at a scale $\mu_{S^0} = {\cal O} (m_{S^0})$, we obtain for the initial condition of the Weinberg operator the result 
\beq \label{eq:wS0}
w (\mu_{S^0}) =  \frac{g_s^3 (\mu_{S^0})}{(4 \pi)^4} \, \frac{2}{m_q^2}  \, {\rm Im} \left (Y_{qq}^2 \right )  \, h (x_{q/S^0}) \,,
\eeq 
with 
\beq \label{eq:dicus1}
\begin{split}
h (x) & = \frac{2 x^2-3 x}{8 \left (4 x-1 \right )^2}-\frac{6
   x^3-5 x^2-x }{4 \left (4 x-1 \right )^3} \, \ln x 
  + \frac{6 x^4-6 x^3+3x^2}{4 \left (4 x-1 \right )^3} \, \phi \left(\frac{1}{4
   x}\right) \,.
\end{split}
\eeq
We note that our  function $h(x_{q/S^0})$  corresponds to $h(m_q, m_{S^0})$ as defined in \cite{Dicus:1989va}.  The function~$\phi$ entering~(\ref{eq:wS0}) stems from the two-loop scalar tadpole with two different masses. The corresponding analytic expression reads \cite{Davydychev:1992mt}
\bea \label{eq:phi}
\phi (u) = \begin{cases} \, 4\hspace{0.25mm} 
  \sqrt{\frac{u}{1-u}} \, \text{Cl}_2 \left( 2 \sin^{-1}
    \left(\sqrt{u}\right)\right) , & u \leq 1 \,, \\[4mm]
  \frac{ \frac{\pi ^2}{3} - \ln ^2(4
    u) +2 \ln ^2\left(\frac{1}{2} 
      \hspace{0.25mm} \left[1-\sqrt{1-\frac{1}{u}} \, \right]\right) -4 \hspace{0.25mm} \text{Li}_2
    \left(\frac{1}{2}  \left[1-\sqrt{1-
          \frac{1}{u}} \, \right]\right) }{ \sqrt{1- \frac{1}{u}}} \, , & u > 1
  \,, \end{cases}
\eea
where $\text{Cl}_2 (u) = {\rm Im} \left [\text{Li}_2 \left (e^{iu} \right )\right ]$ denotes the Clausen function and $\text{Li}_2$ is the usual dilogarithm. In the limit of light or heavy internal quark the function (\ref{eq:dicus1}) can be approximated by the corresponding Taylor expansion
\beq \label{eq:hlimit}
h (x) \simeq \begin{cases} - \displaystyle \frac{x }{4} \left ( \ln x + \frac{3}{2} \right ) \,,  &  x \to 0 \,, \\[4mm] \displaystyle  \frac{1 }{16} \left (1 + \frac{1}{3 x} \right ) \,,  &  x \to \infty \,. \end{cases}
\eeq
We stress that in order to obtain the correct result for $w (\mu_{S_0})$ in the case $x \to 0$, one has to take into account two-loop diagrams  in the effective theory. The corresponding matrix element will cancel the $1/ m_q^2$ dependence in~(\ref{eq:wS0}), resulting in a vanishing initial condition of the Weinberg operator.  Notice however that a non-zero coefficient $w(\mu_H)$ is  induced through operator mixing and threshold corrections. Details on the RG-improved calculation of $w (\mu_H)$ in the case of $x \to 0$ can be found in Appendix B of  \cite{Brod:2013cka}.

The interactions relevant for the case of charged scalar $S^+$ exchange can be written as 
\beq
{\cal L} \supset -\left(  Y_{qq^\prime}  \hspace{0.5mm}  \bar q_L q^\prime_R + Y_{q^\prime q}  \hspace{0.5mm}  \bar q_R \hspace{0.25mm} q^\prime_L 
\right)  S^+ + {\rm h.c.} \, ,
\eeq
where $q$ ($q^\prime$) denotes a up-type (down-type) quark. In this parametrisation the matching correction to the Wilson coefficient of the Weinberg operator takes the form 
\beq \label{eq:wSplus}
w (\mu_{S^+}) =  \frac{g_s^3 (\mu_{S^+})}{(4 \pi)^4} \, \frac{2}{m_q \hspace{0.25mm} m_{q^\prime}} \, {\rm Im} \left ( Y_{q q^\prime} \hspace{0.25mm} Y_{q^\prime q}^\ast \right ) \, h^\prime (x_{q/S^+},x_{q^\prime/S^+}) \,,
\eeq 
where 
\bea \label{eq:chargedhiggs}
\begin{split}
& h^\prime(y,z) = \frac{y^4-y^3 (4 z+5)+y^2 \left(6 z^2+5 z+7\right)-y
   \left(4 z^3-5 z^2+10 z+3\right)+(z-3) (z-1)^2 z}{16
   \left( \left ( 1 - y - z \right)^2 - 4 y z \right)^2} \\[2mm]
   & + \frac{y \left(-3 \left(y^2+3 y-6\right) z^2+\left(y^3+16
   y^2-9 y-8\right) z+(3 y-8) z^3+(y-1)^3-z^4\right)}{4
   \left(  \left ( 1 - y - z \right)^2 - 4 y z  \right)^3} \ln y  \\[2mm]
   & + \frac{3 y^2 z \left(-y^3+y^2 (z+1)+y \left(z^2-4
   z+1\right)-(z-1)^2 (z+1)\right)}{4 \left(  \left ( 1 - y - z \right)^2 - 4 y z \right)^4} \ \psi \left ( \frac{1}{y}, \frac{z}{y} \right ) + \big( y \leftrightarrow z \big ) \,.
\end{split}   
\eea
In \cite{Dicus:1989va} the  loop function corresponding to $h^\prime (x_{q/S^+},x_{q^\prime/S^+})$ is denoted by $h^\prime (m_q, m_{q^\prime}, m_{S^+})$. The function $\psi$ arises from the two-loop scalar tadpole integral involving three different mass scales. It is given by  \cite{Davydychev:1992mt}
\bea \label{eq:psi}
\psi (v,w) = \begin{cases} -2 \hspace{0.25mm} \sqrt{-\lambda^2} \; \, \bigg [ \, \text{Cl}_2 \left ( 2 \cos^{-1} \left [ \frac{-1+v+w}{2\sqrt{vw}} \right ] \right ) + \, \text{Cl}_2 \left ( 2 \cos^{-1} \left [ \frac{1+v-w}{2\sqrt{v}} \right ] \right )  & \\[-1.5mm]  & \; \lambda^2 \leq 0\,, \\[-1.5mm]
\hspace{19mm} + \, \text{Cl}_2 \left ( 2 \cos^{-1} \left [ \frac{1-v+w}{2\sqrt{w}} \right ] \right ) \bigg ] \,,
& \\[4mm]
\lambda \, \Big [ \,  \frac{\pi^2}{3}  - \ln v \ln w & \\[2mm] \phantom{xx} + 2 \ln \left ( \frac{1}{2} \left [ 1 + v - w - \lambda \right ] \right ) \ln \left ( \frac{1}{2} \left [ 1 - v + w - \lambda \right ] \right ) &  \; \lambda^2 > 0 \,, \\[2mm] \phantom{xx}  - 2 \hspace{0.25mm} \text{Li}_2  \left (  \frac{1}{2} \left [ 1 + v - w - \lambda \right ] \right ) - 2 \hspace{0.25mm} \text{Li}_2  \left (  \frac{1}{2} \left [ 1 - v + w - \lambda \right ] \right )  \Big ] \,,   \end{cases}
\eea
with $\lambda \equiv \sqrt{(1-v-w)^2-4vw}$. Note that the result for $ \lambda^2 \leq 0$ ($\lambda^2 > 0$) in (\ref{eq:psi}) was obtained in the region $\sqrt{v} + \sqrt{w} \geq 1$ ($\sqrt{v} + \sqrt{w} \leq 1$). By permutation of the mass parameters it is however straightforward to find the analytic results in the remaining regions. In the limit of infinitesimally small (large) mass $m_q$, the function $h^\prime(y,z)$ behaves like
\beq \label{eq:hyzlimit}
h^\prime(y,z) \simeq \begin{cases} \displaystyle \frac{z^2-3 z}{8 \left (z-1 \right )^2}+\frac{z}{4 \left (z-1 \right )^3} \, \ln z \,, & y \to 0 \,, \\[6mm] 
\displaystyle \frac{1}{8} \left (1 - \frac{1}{z} \right ) \,, & y \to \infty \,. \end{cases}
\eeq
Notice that for  $y \to 0$ one has $h^\prime(y,z) \simeq z/8 \, f_1(z)$ with $f_1$ given in (\ref{eq:f}). The appearance of the loop function $f_1$ is not accidental, since the term in (\ref{eq:wSplus}) proportional to $1/m_q$ has to cancel after including  matrix elements in the effective theory. These complications are avoided if the calculation is performed with $m_q =0$, as done in Section~\ref{sec:indirecttch}. In such a calculation all diagrams in the effective theory lead to scaleless integrals which evaluate to zero in dimensional regularisation. In consequence, the initial condition~$w (\mu_{S^+})$ is then obtained directly from the two-loop graphs in the full theory.

\end{document}